# Controlling magnetism in 2D CrI$_3$ by electrostatic doping

Shengwei Jiang [1,2], Lizhong Li [1,2], Zefang Wang [1,2], Kin Fai Mak [1,2,3]*, and Jie Shan [1,2,3]*

[1] School of Applied and Engineering Physics, Cornell University, Ithaca, New York 14853, USA
[2] Laboratory of Atomic and Solid State Physics, Cornell University, Ithaca, New York 14853, USA
[3] Kavli Institute at Cornell for Nanoscale Science, Ithaca, New York 14853, USA

*E-mails: kinfai.mak@cornell.edu; jie.shan@cornell.edu

**The atomic thickness of two-dimensional (2D) materials provides a unique opportunity to control material properties and engineer new functionalities by electrostatic doping. Electrostatic doping has been demonstrated to tune the electrical [1] and optical [2] properties of 2D materials in a wide range, as well as to drive the electronic phase transitions [3]. The recent discovery of atomically thin magnetic insulators [4,5] has opened up the prospect of electrical control of magnetism and new devices with unprecedented performance [6-8]. Here we demonstrate control of the magnetic properties of monolayer and bilayer CrI$_3$ by electrostatic doping using a dual-gate field-effect device structure. In monolayer CrI$_3$, doping significantly modifies the saturation magnetization, coercive force and Curie temperature, showing strengthened (weakened) magnetic order with hole (electron) doping. Remarkably, in bilayer CrI$_3$ doping drastically changes the interlayer magnetic order, causing a transition from an antiferromagnetic ground state in the pristine form to a ferromagnetic ground state above a critical electron density. The result reveals a strongly doping-dependent interlayer exchange coupling, which enables robust switching of magnetization in bilayer CrI$_3$ by small gate voltages.**

Control of magnetism by electrical means is an attractive approach for magnetic switching applications because of its low-power consumption, high speed and good compatibility with conventional semiconductor industry [6-8]. Electrical control of magnetism has been explored in a variety of magnetic materials including dilute magnetic semiconductors [9,10], ferromagnetic metal thin films [11,12], magneto-electrics [13] and multiferroics [14-16]. The emergence of atomically thin magnetic insulators/semiconductors [4,5] that can be readily integrated into van der Walls heterostructures [17,18] to form field-effect devices has presented a unique and promising system for electrical control of magnetism. A recent experiment has shown electrical control of the magnetic order in bilayer CrI$_3$ based on a linear magneto-electric effect [19] (Supplementary Sect. 3). Such a strategy, however, is limited to non-centrosymmetric materials [6,13-16] magnetically biased near the antiferromagnet - ferromagnet transition [19]. Here we demonstrate effective tuning of the magnetic properties of both monolayer and bilayer CrI$_3$ by a more general approach based on electrostatic doping. Our results provide the basis for future voltage-controlled spintronic and memory devices based on 2D magnetic materials, as well as stimuli for future theoretical and experimental investigations of the microscopic mechanisms of the observed physical phenomenon.

Pristine monolayer CrI$_3$ has been shown a model Ising ferromagnet below ~ 50 K [4]. The magnetic moment carried by the Cr$^{3+}$ ions, which are arranged in a honeycomb lattice structure and octahedrally coordinated by nonmagnetic I$^-$ ions, is aligned in the out-of-plane direction by anisotropic exchange interaction mediated by the I$^-$ ions [20-22]. On the other hand, pristine bilayer CrI$_3$ has been shown an antiferromagnet with antiparallel magnetization from two ferromagnetic (FM) monolayers below a critical temperature of ~ 58 K [4]. The antiferromagnet - ferromagnet transition occurs at a low critical field of ~ 0.6 – 0.7 T at low temperature [4], reflecting the interlayer exchange interaction being weak in the system [23]. The weak interlayer exchange interaction is expected to be susceptible to external perturbations, such as doping, thus providing a unique route for nonmagnetic control of the antiferromagnet - ferromagnet transition as we demonstrate below.

To electrically gate atomically thin CrI$_3$, we fabricated dual-gate field-effect devices using the van der Waals assembly method [18, 24]. Figure 1a shows the schematic side view of our devices, and figure 1b, optical micrograph of two sample devices. CrI$_3$ is encapsulated in either hexagonal boron nitride (hBN) or graphene thin layers to minimize environmental effects. Graphene is also used as contact and gate electrodes, and hBN as gate dielectric. The dual-gate structure allows independent control of the electric field and the doping level, as well as achieving high doping levels in the channel. We will focus on the effect of doping in this study. For nearly symmetric top and back gates the doping level is controlled by the sum of the two gate voltages, which we refer to simply as gate voltage below. The gate-induced doping density in CrI$_3$ was calculated from the gate voltage using a parallel-plate capacitor model with the dielectric constant ($\approx$ 3 [25, 26]) and thickness of the hBN gate dielectric measured independently. We estimate the uncertainty for the doping density to be on the order of 10 % due to uncertainties in the hBN thickness and dielectric constant. Because the electronic density of states in CrI$_3$ is substantially higher than that in graphene due to the flat Cr $d$-bands [22, 27-29], the presence of graphene contacts embedding the CrI$_3$ channel has little influence on the efficiency of doping the magnetic material. Details on the device fabrication and characterization are provided in Methods and Supplementary Sect. 2.

The magnetization of CrI$_3$ was characterized by the magnetic circular dichroism (MCD) at 633 nm using a confocal microscope. The MCD, linearly proportional to sheet magnetization $M$, can be calibrated by assuming that under saturation each Cr$^{3+}$ ion carries a magnetic moment of $3\mu_B$ ($\mu_B$ is the Bohr magneton) in pristine samples [20, 21]. Figure 1c and 1d are the magnetic-field $\mu_0 H$ dependence of the MCD for a monolayer and bilayer device, respectively, at 4 K near zero doping ($\mu_0$ is the vacuum permeability). The results are fully consistent with an earlier report [4]. Namely, monolayer CrI$_3$ is a ferromagnet with a coercive force $\mu_0 H_c$ of ~ 0.13 T, and bilayer CrI$_3$ is an antiferromagnet which turns into a ferromagnet at a spin-flip transition field $\mu_0 H_{sf}$ of ~ 0.6 T. A clear hysteresis is seen for the antiferromagnet - ferromagnet phase transition, indicating its first-order nature. To accurately determine the Curie temperature $T_C$ of monolayer CrI$_3$, we measured the temperature dependence of its magnetic sheet susceptibility $\chi(T) = \frac{\partial M(T)}{\partial H}$ (Fig. 2b). To this end, a conducting ring was fabricated around the sample (left, Fig. 1b) and an ac current at 30 kHz was applied to the ring to generate a small ac magnetic field of ~ 6 Oe in amplitude at the sample. The modulated

MCD was measured by a lock-in amplifier under zero dc magnetic field to yield the sheet susceptibility $\chi$. The value of $T_C$ was extracted by analyzing the behavior of $\chi(T)$ when $T_C$ is approached from above. The solid lines in Fig. 2b are fits to the data using the result for a 2D Ising model: $\chi \propto (T - T_c)^{-\gamma}$ with a critical component γ = 1.75 [30]. See Methods and Supplementary Sect. 1 for more details.

We first examine the effect of electrostatic doping on the magnetic properties of monolayer CrI$_3$. The top and bottom panels of Fig. 2a are the magnetic-field dependence of magnetization *M*(*H*) at three selected gate voltages at 4 K and 50 K, respectively. Significant doping-induced changes in the FM hysteresis loop can be observed. In contrast, the effect of pure electric field on *M*(*H*) is negligible (Supplementary Sect. 3). At 4 K, both the saturation magnetization $M_S$ and the coercive force $H_C$ increase (decrease) with hole (electron) doping while the shape of the hysteresis loop remains approximately unchanged. At 50 K, the shape of *M*(*H*) also changes significantly, indicating a doping-induced change in the Curie temperature. This is fully consistent with the magnetic susceptibility measurement at varying doping levels (Fig. 2b). We summarize the doping effect on magnetic properties of monolayer CrI$_3$ in Fig. 2c. The saturation magnetization, coercive force and Curie temperature all have been normalized to their values at zero gate voltage. For the entire doping range, all three parameters increase (decrease) linearly with hole (electron) doping (solid lines are linear fits to the experimental data), i.e. hole (electron) doping strengthens (weakens) the magnetic order in monolayer CrI$_3$. Significant tuning range up to ~ 75%, 40% and 20% has been achieved for $H_C$, $M_S$, and $T_C$, respectively. The result also shows that among these three parameters, the coercive force can be tuned the most and the Curie temperature the least.

Next we examine the effect of electrostatic doping on bilayer CrI$_3$. Figure 3a shows the magnetic-field dependence of the MCD at several representative gate voltages at 4 K. A small remnant magnetization is present in the antiferromagnetic (AFM) phase due to a built-in electric field in the sample and the corresponding magneto-electric effect. It can be completely removed by applying an opposing electric field through gating (Supplementary Sect. 3). The remnant magnetization has a weak doping dependence (Fig. 4c trace 5). (We therefore did not apply any electric field to cancel it in order to access a wider range of doping densities using the combination of two gates.) This behavior is consistent with the picture of two antiferromagnetically coupled FM monolayers: since the two monolayers are doped nearly equally, the doping-induced change in *M* from two monolayers largely cancels out. For a similar reason, in the FM phase the saturation magnetization, which is the combined magnetization of the two monolayers, follows a linear doping dependence as in monolayer CrI$_3$ (Fig. 4c, trace 1).

The most remarkable observation, however, is the drastic change with doping in the spin-flip transition field $H_{sf}$. The field decreases monotonically with electron density and eventually drops to zero at a critical applied density of ~ 2.6×10$^{13}$ cm$^{-2}$. This indicates that the AFM phase has vanished and the material has turned into a ferromagnet above this critical density! A detailed doping density-magnetic field phase diagram is mapped out in Fig. 3b. The AFM phase shrinks continuously with increasing electron density. The extracted spin-flip field $H_{sf}$ (averaged over forward and backward scans) as a function of applied density is shown in Fig. 3c (right axis). It spans a range of ~ 0.6 - 0.7 T. Doping, therefore, not only changes the properties of constituent monolayers in bilayer CrI$_3$, but also significantly modifies the interlayer exchange coupling. The interlayer exchange constant $J_\perp$ can be estimated from the experimental result by noting that $H_{sf}$ can be viewed as an exchange bias field produced by one of the layers in bilayer CrI$_3$ acting on the other layer [31]. This allows us to write the following expression

$$2J_\perp = \mu_0 M_S (H_{sf} - M_S/2t) \tag{1}$$

by comparing the free energy of the AFM and FM phases at the transition (Supplementary Sect. 4). Here the sign convention is $J_\perp > 0$ for AFM coupling, and $< 0$ for FM coupling; $t \approx 0.7$ nm is the interlayer separation [20, 21]. The second term on the right hand side takes into account the demagnetization energy in the FM phase when $H > H_{sf}$. Using the measured doping dependence for $H_{sf}$ and $M_S$, we obtain the doping dependence of $J_\perp$ in Fig. 3c (left axis). As expected, it decreases monotonically with electron doping and changes sign (i.e. changes from AFM to FM coupling) at a critical applied density of $\sim 2.1 \times 10^{13}$ cm$^{-2}$. The material, however, remains AFM until the doping density reaches $\sim 2.6 \times 10^{13}$ cm$^{-2}$, at which the demagnetization energy is overcome by the interlayer exchange interaction.

The microscopic mechanisms for the observed doping effects on magnetism in 2D CrI$_3$ remain not understood and present an exciting topic for future theoretical studies. Our experimental findings suggest the importance of the doping effect on intralayer Cr-Cr exchange interaction, anisotropy of the exchange interaction, and the interlayer Cr-Cr exchange interaction mediated by I$^-$ ions in few-layer CrI$_3$. (See Supplementary Sect. 5 for more discussions). For instance, the large tuning in the saturation magnetization cannot be explained alone by the doping effect on the electron occupancy of the Cr$^{3+}$ ions (which carry all the magnetic moment in CrI$_3$). Due to the octahedral crystal field and the exchange interaction, the Cr $d$-orbitals are split into spin-polarized $t_{2g}$ and $e_g$ levels [22, 27-29]. Recent *ab initio* calculations show that the energy of the fully filled majority-spin $t_{2g}$ levels is significantly below the valence band edge, which is largely composed of spin-unpolarized I $p$-orbitals [22, 27-29]. Our observation of decreasing $M_S$ with electron doping thus suggests that the sample is unintentionally n-doped and the conduction band is composed of minority-spin Cr $t_{2g}$ levels. This is consistent with a recent *ab initio* study [22] but disagrees with earlier *ab initio* studies [27-29], highlighting the need for more accurate band structure calculations. Furthermore, based on this picture of band filling, only up to a $\sim 3\%$ change in $M_S$ is expected in the monolayer (the maximum doping density range achieved in our experiment $\sim 5.3 \times 10^{13}$ cm$^{-2}$ corresponds to $\sim 0.1$ electron per Cr$^{3+}$ ion). This is nearly an order of magnitude smaller than the observed value of 40% (Fig. 2c), suggesting other mechanisms also in play.

Finally, we explore switching of magnetization in 2D CrI$_3$ by electrostatic doping. This is highly relevant to the technologically important area of voltage control of magnetism [6-8]. Figure 4a shows an attempt with a monolayer device. The device was first prepared in the magnetization "up" state by applying a magnetic field at 0.8 T. The gate voltage was then swept while the device was biased right below its coercive force at -0.12 T. As the gate voltage was scanned from negative to positive values, the device was switched from the "up" to the "down" state. The switching is enabled by the doping-dependent coercive force (Fig. 2c). It occurs when the coercive force drops below the bias magnetic field. However, it is one-time only switching as the device stayed in the "down" state afterwards regardless of the gate voltage because the "down" state has a lower energy than the "up" state in monolayer CrI$_3$ under a negative bias magnetic field. The system cannot overcome the energy barrier to go back to the "up" state. Such one-time switching of ferromagnets through tuning $H_C$ electrically has been observed in other magnetic systems [10].

In contrast, the magnetization in bilayer CrI$_3$ can be repeatedly switched between the "up" or "down" state (FM state) and the "zero" state (AFM state). As shown in Fig. 4b, a bilayer device was first prepared in the "up" state by applying a magnetic field at 1 T. The gate voltage was then swept while a bias field at 1, 0.6, 0.5, 0.3 and 0 T was applied, corresponding to trace 1 through 5 in Fig. 4c, respectively. Complete and repeatable switching between the "up" and "zero" states has been achieved in this device for the three intermediate bias fields. The switching is enabled by the doping-dependent $J_\perp$, which changes sign with doping and shifts the global energy minimum of the system between the AFM and FM states. Compared to earlier studies based on the magneto-electric effect to tune the exchange bias for electrical switching of magnetic moments [13-15, 19], switching of magnetization by electrostatic doping here is more efficient and is applicable for a much wider range of external bias magnetic field because of the much stronger doping dependence of the exchange bias field $H_{sf}$ (Fig. 3c). Our findings have thus identified an efficient and versatile approach based on electrostatic doping to modulate magnetization in 2D magnets, and bilayer CrI$_3$ as a particularly promising system for voltage-controlled magnetic switching applications.

**Methods**

**Device fabrication.** Van der Waals heterostructures of atomically thin CrI$_3$, hexagonal boron nitride (hBN) and graphene were fabricated by the layer-by-layer dry transfer method [24, 26]. Images of sample devices are shown in Fig. 1b. First, atomically thin samples were mechanically exfoliated from their bulk crystals (HQ Graphene) onto silicon substrates covered by a 300-nm thermal oxide layer. The thickness of thin flakes was initially estimated from their optical reflectance contrast on silicon substrates and later verified by the atomic force microscopy measurements. Monolayer and bilayer CrI$_3$ were further confirmed by the magnetization measurement under a varying out-of-plane magnetic field. The typical thickness of hBN gate dielectrics was ~ 20 nm, and the typical thickness of graphene was ~ 3 layers. The selected thin flakes were then picked up one-by-one by a stamp consisting of a thin layer of polycarbonate (PC) on polydimethylsiloxane (PDMS). The complete stack was then deposited onto substrates with pre-patterned Au electrodes. The residual PC on the device surface was dissolved in chloroform before measurements. Since atomically thin CrI$_3$ is extremely air sensitive, it was handled inside a glovebox with less than one-part-per-million oxygen and moisture and was removed from the glovebox only after being fully encapsulated.

**Electrical characterization.** The sample conductance in the out-of-plane direction was measured by biasing CrI$_3$ using the graphene source and drain electrodes. Due to the high sample resistance/small conductance, the bias voltage was modulated at a low frequency of ~ 17 Hz and the resultant current was measured with a lock-in amplifier. The sample conductance was recorded as a function of doping density by varying the (sum) gate voltage while the electric field generated by the top and back gate voltage cancels each other at the sample (Fig. 1c). See Supplementary Sect. 2 for more details.

**Magnetization measurement.** The sample magnetization was characterized by the magnetic circular dichroism (MCD) microscopy in an Attocube closed-cycle cryostat (attoDry1000) down to 4 K and up to 1.5 Tesla in the out-of-plane direction. The MCD microscopy was performed using a HeNe laser at 633 nm with an optical power of ~ 5 µW. An objective of numerical aperture (= 0.8) was used to focus the excitation beam to a sub-micron spot size on the devices. The reflected light was collected by the same objective and detected by a photodiode. The helicity of the optical excitation was modulated between left and right by a photoelastic modulator (PEM) at 50.1 kHz. The MCD was determined as the ratio of the ac (measured by a lock-in amplifier) and dc component (measured by a multimeter) of the reflected light intensity.

The MCD was converted to sheet magnetization $M$ by assuming that it is linearly proportional to $M$ and the saturation magnetization in pristine samples is $M_S = 0.137$ mA per layer. The saturation magnetization was obtained by assuming each $Cr^{3+}$ cation carry a magnetic moment of $3\mu_B$ [20, 21] and evaluating the density of $Cr^{3+}$ ions from the crystallographic data of bulk $CrI_3$ (space group $R\bar{3}$ with unit cell parameters of $a = 6.867$ Å, $b = 6.867$ Å, $c = 19.807$ Å and $\beta = 90°$) [20, 21].

The coercive force $H_C$ for monolayer $CrI_3$ and the spin-flip transition field $H_{sf}$ for bilayer $CrI_3$ are defined as the field corresponding to the steepest change in the sheet magnetization $M(H)$. They were determined from the peak position of d$M$/d$H$, which was numerically calculated from the measured $M(H)$. The spin-flip transition is relatively broad. We recorded the full-width-at-half-maximum (FWHM) of the peaks as the transition width (vertical bars in Fig. 3c).

**Magnetic susceptibility measurement.** The ac magnetic susceptibility $\chi$ of 2D $CrI_3$ was measured using an excitation ring around the sample (Fig. 1b, left). An ac current of 50 mA at 30 kHz was applied to the ring to generate a small oscillatory magnetic field of ~ 6 Oe in amplitude at the sample. The MCD under zero dc magnetic field was measured by a lock-in amplifier to yield $\chi$. For each doping level, the value of $\chi$ was measured as a function of temperature. To this end, the sample temperature was first raised to 90 K (well above $T_C$), and the measurement was followed during cooling. The value of $T_C$ can be obtained by fitting the temperature dependence of $\chi$ above $T_C$ using the result for a 2D Ising model as described in the main text. The extracted value of $T_C$ (dashed lines in Fig. 2b) also agrees well with the peak position of $\chi(T)$.

**Figures and figure captions**

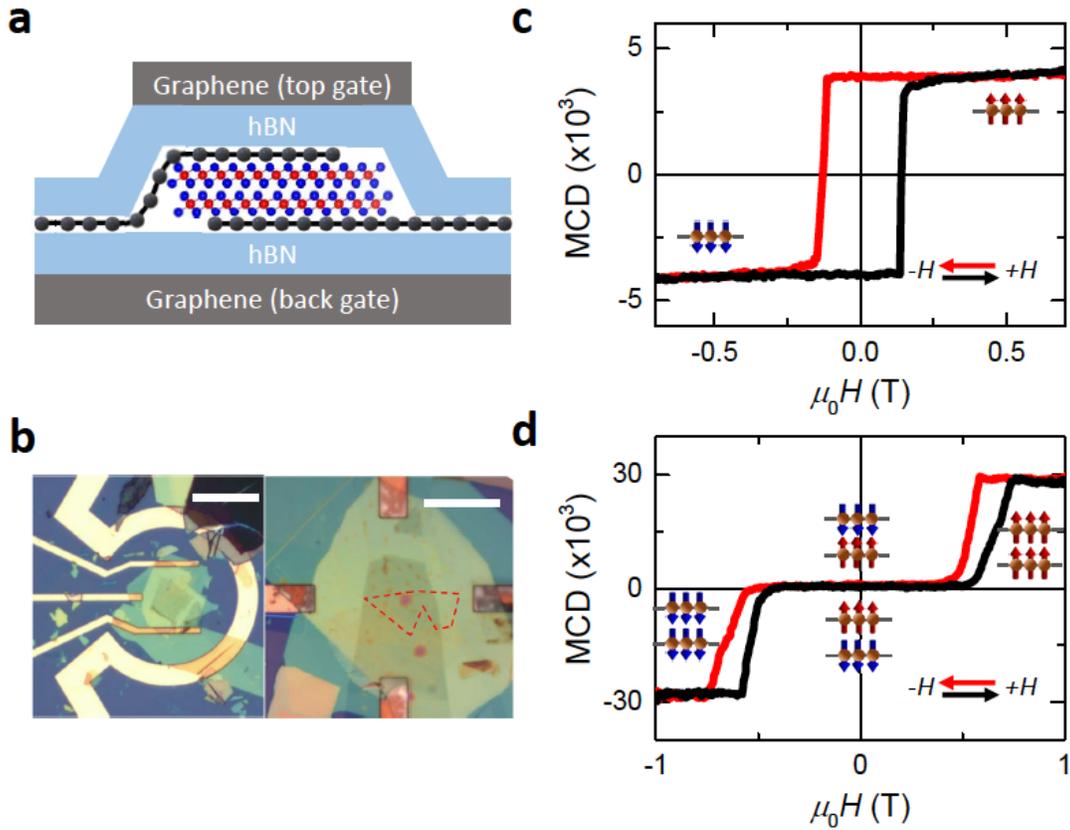

**Figure 1 | 2D CrI$_3$ field-effect devices. a,** Schematic side view of a dual-gate bilayer CrI$_3$ field-effect device. Bilayer CrI$_3$ is encapsulated in few-layer graphene, which also serves as source and drain electrodes for out-of-plane transport measurements. The top and back gates are made of hBN and graphene. **b,** Optical micrograph of two sample devices. The scale bars are 50 and 20 $\mu$m in the left and right panel, respectively. The metallic ring structure (left panel) was used to create an ac magnetic field for the susceptibility measurement for monolayer CrI$_3$. The red dashed line marks the boundary of a bilayer sample in the right panel. **c, d,** MCD versus magnetic field for monolayer (**c**) and bilayer CrI$_3$ (**d**) near zero doping at 4 K. The insets illustrate the magnetic states corresponding to various ranges of magnetic field.

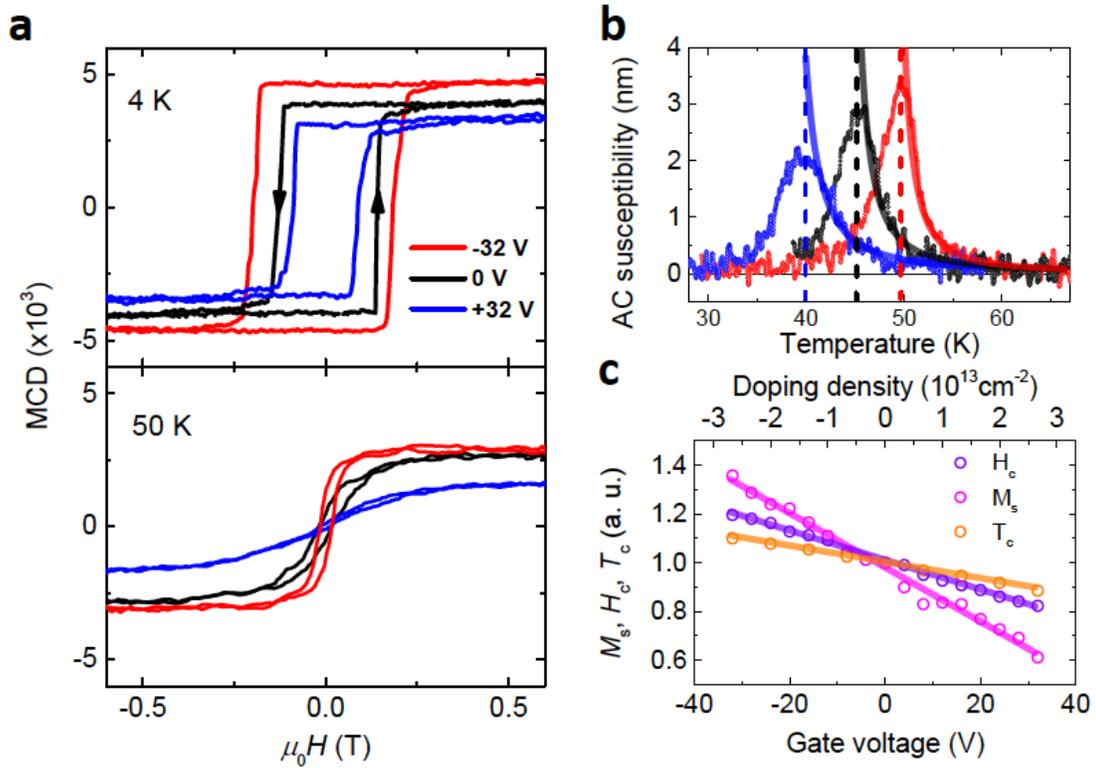

**Figure 2 | Controlling ferromagnetism in monolayer CrI$_3$ by doping. a,** MCD signal versus magnetic field at three representative doping levels at 4 K (top panel) and 50 K (bottom panel). **b,** AC susceptibility (symbols) as a function of temperature measured at the same three doping levels as in **a**. Solid lines are fits to the experimental data above the Curie temperature using the critical dependence of a 2D Ising model as described in the text. The dashed lines indicate the extracted Curie temperatures. **c,** Coercive force (purple), saturation magnetization (pink) (both at 4 K), and Curie temperature (orange) normalized by their values at zero gate voltage as a function of gate voltage (bottom axis) and applied doping density *n* (top axis) with positive (negative) values for applied electron (hole) density. The lines are linear fits to the data.

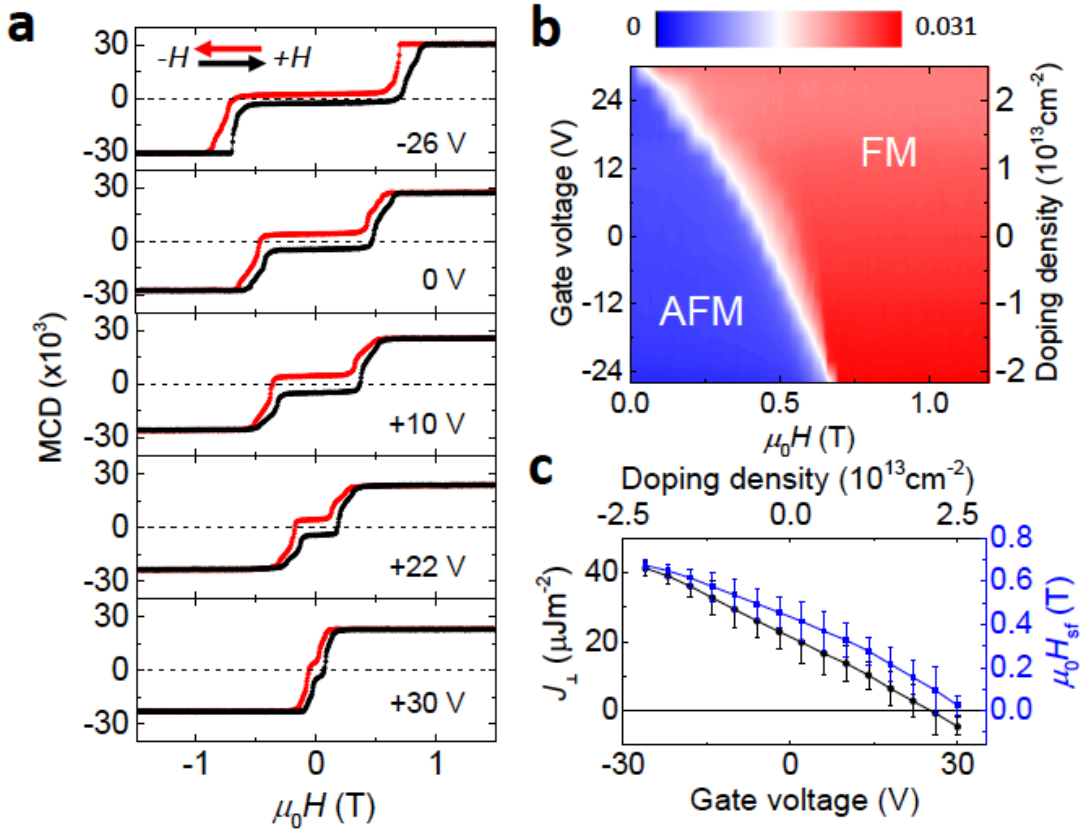

**Figure 3 | Doping-controlled interlayer magnetism in bilayer CrI$_3$. a,** MCD signal versus magnetic field at 4 K at representative gate voltages (Device #2). The finite remnant magnetization in the AFM phase is caused by a built-in interlayer electric field (see main text for details). The spin-flip transition field decreases monotonically with electron doping. **b,** Doping density - magnetic field phase diagram at 4 K. The gate voltage is given in the left axis and the applied doping density in the right axis. The FM and AFM phase correspond to the region of high and low MCD signal, respectively. **c,** Interlayer exchange constant $J_\perp$ (black, right axis) and spin-flip transition field $H_{sf}$ (blue, left axis) as a function of gate voltage (bottom axis) and applied doping density (top axis). The spin-flip transition width determined from the $M(H)$ measurement was shown as error bars for $H_{sf}$, and the errors for $J_\perp$ were propagated from that for $H_{sf}$ using Eqn. (1). The material turns into a ferromagnet at a doping density of $\sim 2.6\times10^{13}$ cm$^{-2}$.

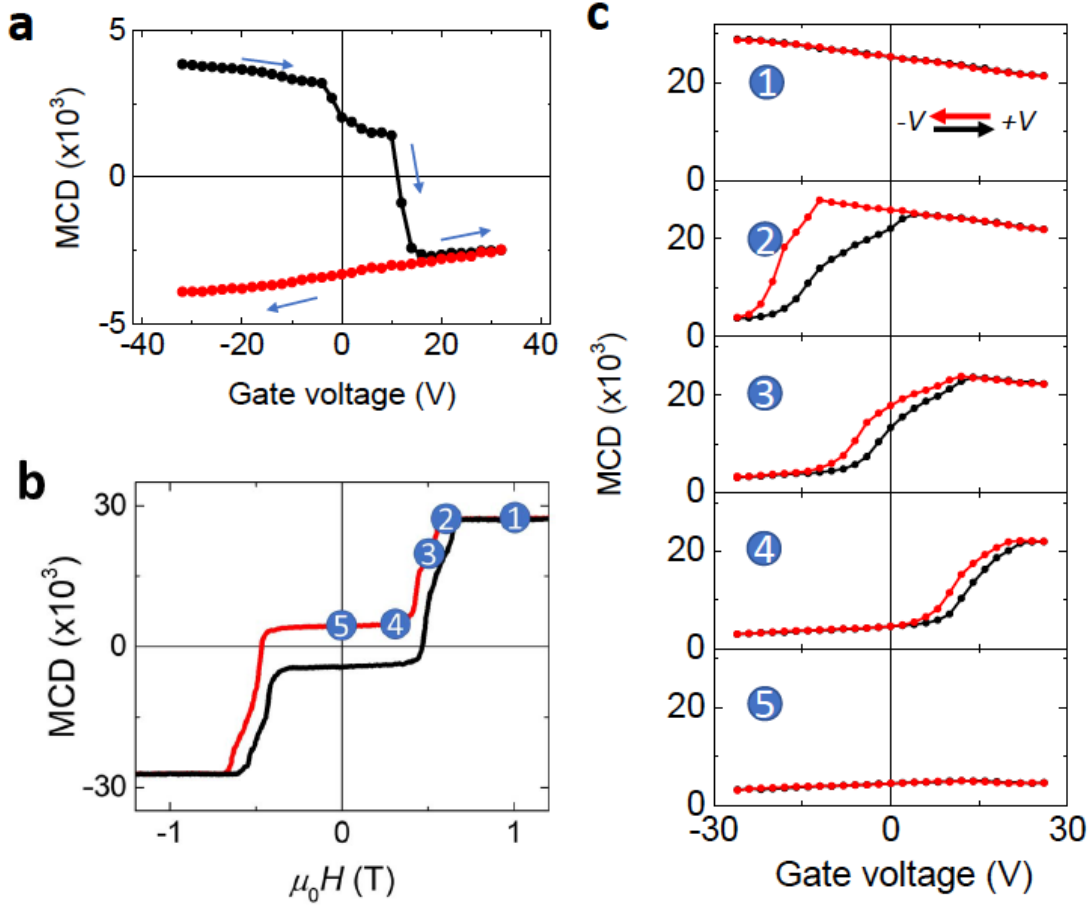

**Figure 4 | Switching of magnetism in 2D CrI$_3$ by electrostatic doping. a,** Gate-voltage control of MCD of monolayer CrI$_3$ at 4 K (Device #1). The sample was prepared in the "up" state by a magnetic field at 0.8 T and then biased at − 0.12 T during the gate voltage sweeps. The magnetization is switched from the "up" to the "down" state at ~ 10 V when the gate voltage sweeps from negative to positive values. It stays in the "down" state in subsequent sweeps of the gate voltage. **b,** The MCD of bilayer CrI$_3$ (device #2) versus magnetic field under zero gate voltage at 4 K. **c,** Gate-voltage control of MCD of bilayer CrI$_3$ at 4 K. The sample was prepared in the "up" state by a magnetic field at 1 T and then biased at 1, 0.6, 0.5, 0.3 and 0 T corresponding to trace 1 through 5, respectively. Black and red curves denote forward and backward sweep directions. Under bias field of 1 T (trace 1) and 0 T (trace 5), the magnetization stays in the "up" (FM) and "zero" (AFM) state, respectively. Under intermediate bias fields (trace 2, 3 and 4), repeatable gate-voltage switching between the "up" and "zero" state has been achieved.